# Habitability of M dwarfs is a problem for the traditional SETI

Milan M. Ćirković[1]

Branislav Vukotić

*Astronomical Observatory of Belgrade, Volgina 7,*
*11000 Belgrade, Serbia*

**Abstract**. We consider some implications of the much-discussed circumstellar habitable zones around M-dwarf stars for the conventionally understood radio SETI. We argue that the flaring nature of these stars would further adversely impact local development of radio communication and that, therefore, their circumstellar habitable zones should be preferentially studied by other methods. This is a clear example how diversity of astrobiological habitats is introducing contingency into the cultural evolution, thus undermining the universality of cultural convergence as one of the major premises of the traditional SETI. This is yet another example of how specifics of the physical environment strongly shape cultural evolution taken in the broadest, most inclusive sense.

## 1. Introduction

There has been a dramatic surge of interest in planetary habitability following discovery of large number of small, Earth-sized planets in exoplanetary systems of red dwarf stars such as Proxima Centauri [1], Kepler 42 [2, 3], or TRAPPIST-1 [4]. Gradually emerging consensus suggests that, just as M-dwarfs dominate with more than 70% in any representative sample of the Main Sequence stars [5], planets around M-dwarfs dominate the overall tally of planets in

---

the Galaxy. The fact that M-dwarfs might not be easy to find and study in biased, usually flux-limited surveys (cf. Figure 1), should not occlude this ground truth: any general statement concerning habitability, biosignatures, technosignatures, etc. should be conditionalized on the statistical predominance of red dwarf stars.

This includes the halo population of very old red dwarfs as well [6]. Obviously, consequences for astrobiology are dramatic: habitability of exoplanets and exomoons around red dwarfs has become one of the most discussed issues in recent astrobiological research [7–15]. While all results are necessarily very tentative as of this writing, there is a growing trend of accepting planets around red dwarf as legitimate part of the overall set of habitable planets in the Galaxy.

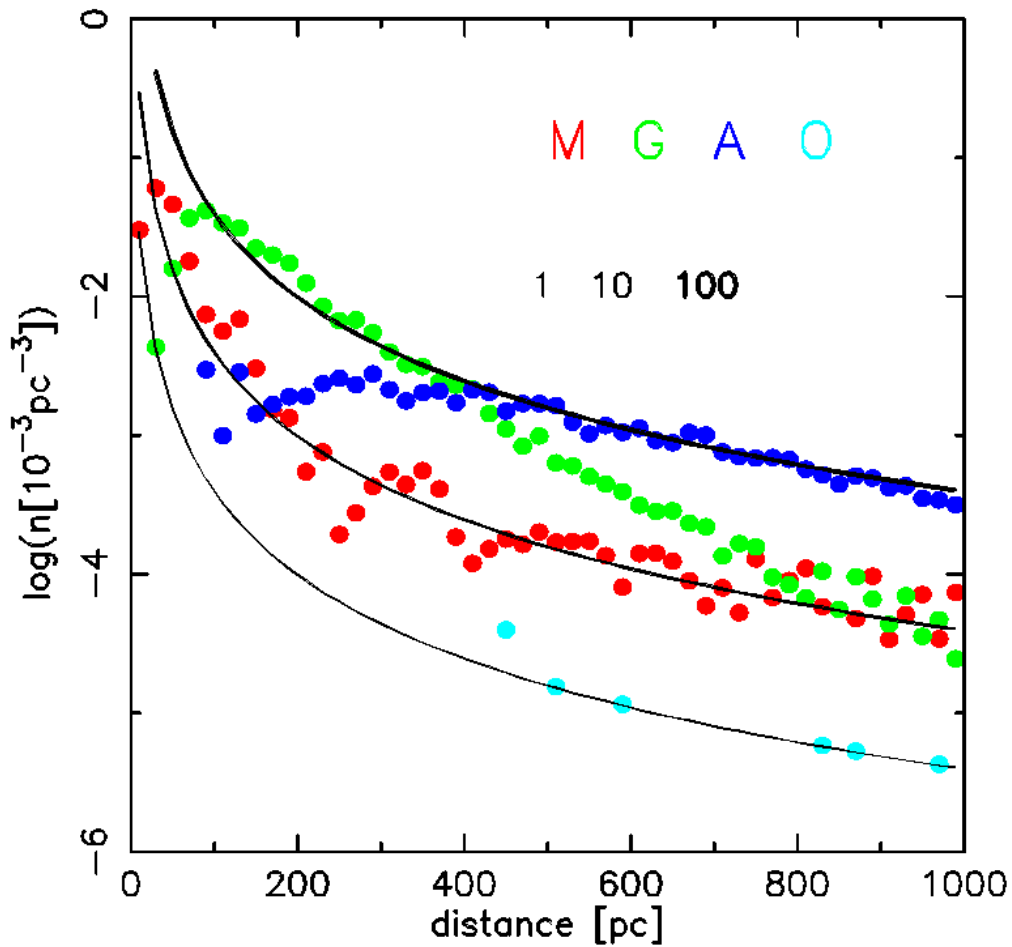

**Figure 1.** The average density of stars with distance. Stars are counted in spherical shells of 20 pc width. The spectral types are color-coded as designated while the black lines of different width represent the density for the number of stars at the shell at the given distance, as designated in the plot. The data was queried per spectral type from the SIMBAD database on October 12, 2019.

Even modern studies on the rate of habitable planetary formation throughout the Milky Way history are nowadays led to separate red dwarf planetary systems from the rest of the Main Sequence stars [16, 17]. Including the former leads to significantly different values of such crucial quantities as the median age of habitable planets in the Milky Way. In fact, the emerging position seems to be that an *average* habitable planet should be located around a red dwarf star, prompting a natural question why are we – as per an anthropic selection effect – not located around one as well [18]. Even if we conclude that the observers of our kind are not vanishingly rare in the overall astrobiological landscape, it is still worthwhile to inquire whether our specific SETI methodologies are, in one way or another, inconspicuously rooted in peculiarities of this particular reference class of observers. In other words, we need to ask ourselves to what extent the detectability of other types of observers is a consequence of the systemic properties of their physical environment – as opposed to being product of the contingencies and happenstance of the entire (biological + cultural) history.

Consider the theoretical assumptions of our own SETI projects. Since the early days of the "Founding Fathers" in 1950s and 1960s, most practical activities in the field have relied on radio waves as presumed carrier of extraterrestrial signals. The precepts of Cocconi and Morrison [19] held firm in this regard, and in fact any suggestion of an alternative SETI, theoretical or empirical, has been often downplayed as fantastic or simply ignored. The key resource database is https://technosearch.seti.org, maintained by the SETI Institute. That site is an attempt to catalog all published SETI searches since Project Ozma in 1960. It is organized into three different lists: Radio, Optical, and Archival; "optical" list includes wavelengths from γ-rays through the far-infrared, and "archival" list contains searches that have used previously published data to search for technological artefacts. While some early examples are present, in particular theoretical analyses of advanced technology by Dyson and of optical SETI by Townes and Schwartz in early 1960s, it has been only since the turn of the century (partially for technological reasons) that more serious non-radio searches have been performed. And the atmosphere has changed to the effect that in recent years (roughly since 2010) that the alternative approaches, such as extragalactic (e.g., 20-22) or Dysonian SETI [23] have received more mainstream acceptance. Of course, more work in history of science is necessary in order to elaborate upon this development; the present manuscript has no such aspirations. Instead, we wish to focus upon one particular aspect in which our conventional technological practices in the SETI field are in tension with the ubiquitous evolutionary contingency.

This issue is by no means too specific or parochial: it touches upon the key topic of *cultural convergence* upon which all SETI projects are based (even if it is too often only tacitly admitted by the research community). Following the original argument of Simpson [24], the great geographer and evolutionist Jared Diamond has argued that particular features of human culture, such as radio, are likely to be contingent within the context of cultural evolution [25]. Radio is just a particularly prominent example of direct interest to SETI, but the reasoning we need to reconsider here applies equally to many other aspects of technological evolution.

In this note, we explicitly do *not* wish to enter the complex debate about habitability of planets around M-dwarfs and whether they could support simple life only, or both simple and complex life, etc. Instead, we wish to discuss the following related questions: *if* planets around red dwarfs are indeed habitable for complex life and intelligence, should we then stick to the conventional orthodoxy of radio searches in their cases? Are there clear physical reasons to adopt different, more efficient strategy for such, sufficiently different habitable locales? If the answer to the latter question is in the affirmative, then it is of crucial importance to consider what other anthropocentric cultural baggage [26, 27] we have unjustifiably smuggled in while conceiving and planning our SETI programs.

## 2. Contingent nature of radio communication

Why, exactly, should we assume radio communication as typical in a wider astrobiological context? One of the most influential contemporary intellectuals, evolutionary biologist and historian Jared Diamond has argued forcefully against our SETI efforts on the basis of their blithely assumed convergence. He paraphrases this assumption as follows:[2]

> *If you expose millions of species for millions of years to similar selective forces, of course you can expect similar solutions to emerge time and time again. We know that convergence is very common among the species on Earth, and, by the same reasoning, convergence should also exist between Earth's species and those elsewhere. Hence, although radio communication has evolved here only once so far, considerations of convergent evolution lead us to expect its evolution on some other planets as well.*

Note that Diamond does not only ascribes to the SETI proponents the core idea of continuity of cultural and biological evolution – he does accept it himself in order to show that it is *insufficient* for supporting the SETI position. The essence of his counter-argument rests upon the example of woodpecking. It is an empirical fact that woodpeckers are an extremely well-adapted species, whose complex adaptations enable them to fill a stable ecological niche with significant success. Which further leads him to conclude (p. 162):

> *Consider what biology might have taught us about the inevitability of radio evolution on Earth. If radio building were like woodpecking, some species might have evolved certain elements of the package or evolved them in inefficient form, although only one species managed to evolve the complete package. For instance, we might have found today that turkeys build radio transmitters but no receivers, while kangaroos build receivers but no transmitters. The fossil record might have shown dozens of now-extinct animals experimenting over the last half billion years with metallurgy and increasingly complex electronic circuits, leading to electric toasters in the Triassic, battery-operated rat traps in the Oligocene and, finally, radios in the Holocene.*

---

[2] Ref. 28, p. 158.

A detailed criticism of logical and empirical aspects of Diamond's argument is given in Ćirković [29]. Here we wish to bring attention to another, usually disregarded aspect of this evolutionary argumentation, namely that even if we fully accept the continuity between biological and cultural evolution, as well as high adaptive value of radio communication on Earth, this is still insufficient for the strong convergence conclusion (cf. Ref. 30). In particular, there may be wide classes of habitats in which radio communication has much lower adaptive value, or none at all.

We need explicitly *astro*biological evolution to account for selection pressures at the very high levels of joint (biological + cultural) evolution. In a way, this has been recognized; one of the early SETI pioneers, Michael Papagiannis noticed back in 1984 that the very limits of sustainability might act as a selection pressure, filtering out profligate and wasteful civilizations [31]. This study is one of the very first in which the continuity of biological and cultural evolution was emphasized in its natural, astrobiological context. Obviously, astrophysical distribution of resources dictates the intensity of this kind of selection pressure.

We should expect other abiotic selection pressures, one of them being adequately generalized space weather. The conventional definition of space weather is pragmatic and anthropocentric (e.g., The US National Space Weather Strategy and Action Plan 2019[3]):

> *Space weather comprises a set of naturally occurring phenomena that have the potential to adversely affect critical functions, assets, and operations in space and on Earth. Extreme space weather events can degrade or damage critical infrastructures, which may result in direct or cascading failures across key services such as electric power, communications, water supply, healthcare, and transportation. The Sun can create space weather events that have the potential to substantially disrupt or damage critical functions, assets, and operations in space and on Earth, including critical infrastructure and technology systems.*

---

[3] https://www.whitehouse.gov/wp-content/uploads/2019/03/National-Space-Weather-Strategy-and-Action-Plan-2019.pdf, last accessed September 25, 2019.

Clearly, this statement could and should be generalized to include stars other than the Sun and their habitable and potentially inhabited planets. Space weather generalized in such a way is necessarily part of the abiotic selection pressure acting on the set of technological civilizations in the Galaxy.

It is very well known that space weather influences many aspects of contemporary human civilization on Earth, such as satellites [32], high-latitude electric power grids [33], commercial airplane traffic [34], or human radio communications themselves (e.g., 35-37). Extreme space weather events, like the Carrington event of 1859, present an acknowledged global risk for modern civilization accustomed to electronics and long-range communications [36]. Space weather on and near Earth, however, is *extremely nice and clement* in comparison to other astrophysical locales, and notably in comparison the environment around flaring and magnetically active M-dwarfs. Both persistent and intermittent activity of such stars, coupled with the necessarily small distance of planets within their circumstellar habitable zones, create space weather conditions which are orders of magnitude more extreme than anything that could ever be encountered on and near Earth. Consequent impact on technological systems of our cosmic peers – and targets to be chosen in SETI surveys – should be correspondingly larger and lead to diverging outcomes, when compared with our situation. Taking into account the abundance of M-dwarfs and planets around them, this effect might be dominant overall and seriously jeopardize our conventional SETI strategies.

### 3. M dwarf flares as sources of radio noises

That M-dwarf flares emit substantial electromagnetic noise has been known since the early days of radio astronomy. Direct detection of cool main-sequence stars in the radio domain beyond 10-20 pc is still very difficult, however. Radio emissivity is highest in the very early stages of stellar evolution, near ZAMS, although the ages of M dwarfs are very poorly constrained [38]. Majority of M dwarfs belong to the magnetically active category (160 of 238 in the sample of Ref. 39). These dMe stars possess magnetic fields 10-100 times the strength of Solar magnetic field, and their frequent intensive flares are accompanied by coronal mass ejections and formation of complex persistent magnetospheric structures as large as 10 $R_*$. If the radius of a typical M dwarf is similar to the radius of Proxima, about $1.07 \times 10^5$ km, scales of these magnetic structures are large enough to partially overlap with sizes of the circumstellar habitable zone. Space weather on worlds located in these habitable zones is necessarily extremely complex, and orders of magnitude more intense and inclement than in the case of Earth.

Observations of Proxima by Davenport et al. [40] show that "normal" flares of energies about $10^{28}$ ergs occur on the average 63 times *a day*, while "superflares" at $10^{33}$ ergs or more occur on the average 8 times per year. While flares like these could be major destructive influences on the biospheres of planets around red dwarfs and could possibly cause mass extinction episodes even on Earth, which is 1 AU distant from a quiescent star like the Sun [41], we neglect this more general aspect of flaring here, concentrating on the effects of space weather on presumed technological civilizations. Segura et al. [10] investigate effects of a strong flare of April 12, 1985 on red dwarf AD Leo by simulating the effects from both UV radiation and protons on the atmospheric chemistry of a hypothetical, Earth-like planet located within red dwarf's habitable zone. Of particular concern is the decrease in ozone column density as a consequence of high-energy photons and protons. The conclusions of these authors are, however, optimistic in the sense that the impact of strong flares on the atmospheric ozone is quite modest and unlikely to seriously damage a hypothetical biosphere. Somewhat different results have been obtained by Howard et al. [42] for the Proxima 2016 superflare, their models indicating rather strong ozone depletion at Proxima b, with atmospheric parameters taken to be Earth's. In addition, there are other possibly adverse effects, such as desiccation of planetary water [41]. Irrespectively of this controversy, if we suppose that at least some biospheres around M dwarfs can evolve technological civilizations, they would still face space weather incompatible with the extensive use of radio communications familiar from our experience.

Solar flares are known to cause so-called radio fading events at HF (3–30 MHz) range, characterized by decrease in signal amplitude due to increased ionospheric electron density. Such events on Earth can last up to several hours and cause dramatic increase in the noise in HF bands, including the complete loss of signal [35]. These effects are likely to be magnified by at least two orders of magnitude or more for hypothetical inhabitants of planets around M dwarfs. This is strongly dependent upon the unknown planetary magnetic fields, although it must be noted that planets in habitable zones of red dwarfs are likely to be slow rotators – tidally locked or in Mercury-like resonances – with weak magnetic fields [11].

Several recent studies have measured the intensity of radio bursts of M-dwarf stars (e.g., 43) related to flaring. Many long bursts, of the duration of hour or several hours are detected with peak flux densities of 10–200 mJy in the frequency interval 1–6 GHz, which is exactly the one of interest for the traditional SETI (since it includes the "water hole" range). Frequency drifts also occur in almost all bursts, on variety of timescales; this would make maintenance of reliable radio communication even more difficult. Similarly, the March 24, 2017, radio flare of Proxima reached peak intensity of 100 mJy, about three orders of magnitude higher than the persistent emission in the quiescent state [44]. To further put it into perspective, this peak intensity is about an order of magnitude greater than the typical source flux at 21cm in the VLA SETI observation of Gray and Marvel [45], and 50–100 times the intensity of typical noise in radio maps at 6.25 GHz.

The Arecibo message was transmitted at a power of 1 MW. Assuming a transmitting bandwidth of 10 Hz this translates to radio flux of 10 µJy. With the sensitivity of the upcoming state-of-the-art modern radio telescopes reaching the sub-µJy scales, this message would be detected at distances in excess of 10 pc. However, at a Proxima-like distance of ~1 pc the signal strength would be of the order of ~1 mJy which is some two orders of magnitude smaller than the strength of Proxima Centauri radio flares. The Arecibo message broadcast was less than three minutes in duration, which would make an analogue likely to overlap with periods of high stellar activity. For the detected M dwarfs' radio bursts, Villadsen & Hallinan (Ref. 43, Table 3) report peak flux uncertainties ranging from ~1% to ~40%. All this puts the traditional SETI attempts, such as the Arecibo message, in the noise region of the typical stellar activity.

These comparisons are, of course, of limited value, since it is highly improbable that – taking into account the local space weather – large installations such as the Arecibo telescope would have been built by a hypothetical Proximan civilization in the first place. It seems only reasonable to assume that such a civilization would tend to use other methods of long-range communication both for internal purposes and for hypothetical interstellar communication. Such methods may be those already conceived by present-day humans (e.g., Ref. 46), like gamma-rays or inscribed matter packages, or hitherto unimagined "wild cards".

## 4. Discussion

The idea that radio communication is essentially convergent – that it will emerge independently in the course of (biological + cultural) evolution in many Galactic biospheres – is of key importance for the traditional SETI. Ever since Tesla and Marconi invented radio, via Cocconi and Morrison [19] and the OZMA Project, to the *Breakthrough* Initiatives of today, radio communication has been seen as the most efficient way of transmitting information over long, incl. interstellar distances. The traditional SETI is based on the idea that radio communication is a rule, and that the other ways of communicating like optical lasers, gamma-rays, neutrinos, gravitational waves, or inscribed matter artefacts, if they exist somewhere at all are contingent exceptions, happenstance. Some of the limitations of this approach have been known for quite some time [47]. However, this idea emerged in the context of human history and human technological evolution, predicated upon the relative radio quiescence of our Sun and very clement space weather for most of the time since *homo sapiens* developed technological civilization. These physical circumstances should not be taken for granted at all – and especially should not be taken for granted now that we know much about the spatiotemporal distribution of habitable planets in the Galaxy. In the case of M dwarfs, we may have, ironically, a kind of see-saw effect: while strong magnetic fields may indeed help habitability of planets around such stars, they might simultaneously suppress the kind of technology, namely radio, which could make the emerging technological civilizations there detectable.

Copernican reasoning would suggest that since the one example of a known technological civilization orbits a G-type star, we should – lacking deeper empirical knowledge – expect that others will as well. If M-dwarfs dominate in the set of Galactic habitable planets and if habitable planets around M-dwarfs dominate in the subset of habitable planets evolving intelligent life, this assumption used in the traditional SETI may be unfounded, as described above. A straightforward consequence would be the corresponding methodological shift. Radio communications in the vicinity of M-dwarfs would be much less efficient, less reliable, and more expensive than what has been experienced in human history, our being located around a G-type star. Thus, technological civilizations arising around M-dwarfs would be pressured into developing other mentioned ways of first planetary and later interplanetary and interstellar communication.[4] In other words, the pathway of technological evolution of such intelligent communities will likely diverge from our pathway in several respects. One of these is exactly the key for detection and/or communication over interstellar distances. Hence, our SETI efforts should be modified accordingly. If our present view is correct, we can predict the lack of any SETI-interesting radio sources from the vicinity of red dwarfs, even if it possesses planets in its circumstellar habitable zone [48].

There is a wider and more important philosophical and methodological consideration behind this and similar practical issues: the emancipation from anthropocentrism and fully endorsing the diversity of habitats in the overall astrobiological landscape [27]. It could be regarded as analogous to the process of Bayesian learning: in the beginning we knew very little about physical conditions of other Galactic habitats and as we learn more and more, we *have to* modify and update our optimal strategies. There has been strong recalcitrance to such process in the circles of the traditional SETI. One unfortunate side effect of this recalcitrance is exactly that it strengthens the hand of those radical sceptics which would gut the entire SETI enterprise, as is the case with Jared Diamond. And it is important to notice that there are also distinguished astrobiologists arguing for at best skepticism regarding SETI at least in part on the basis of evolutionary contingency [49, 50].

---

[4] In the specific case of M-dwarfs, UV and soft X-ray communications would be intermittently suppressed as well by their XUV flaring and coronal ejections; cf. [10].

Diamond criticized the *whole* of SETI on this basis; while it can be shown that his general argument fails and that evolutionary convergence does indeed leave the doors open for cautious optimism in regard to SETI [25, 29], there are *particular* SETI issues where Diamondian scepticism is more justified. The issue of hypothetical technological civilizations evolving around M dwarfs in light of the traditional radio SETI is one such example. In light of the above it is unlikely that radio will be a convergent feature of the cultural evolution of such civilizations – if they arise in the first place, of course. Even if they subsequently develop satisfactory shielding from flares (cf. Ref. 51) or colonize different planetary systems with calmer space weather [52], it is quite likely that cultural quirks will persist, be sufficiently entrenched, and suppress, *ceteris paribus*, relative frequency of the radio usage, as compared to the other methods of long-range communication (such as those listed in Ref. 46). Therefore, we need to do the same while targeting – extremely numerous – planetary systems of M-dwarfs in our SETI surveys. In the same time, we need more systematic analysis of our anthropocentric quirks and peculiarities in order to correct for multiple biases of our historically contingent culture and formulate truly objective SETI programs. Hence, the current issue represents an excellent example of the urgent need to overcome the disciplinary barriers and to formulate true multidisciplinary research programs, in which our insights in astrophysics, astrobiology, and SETI studies could be fruitfully joined with those of the studies of culture, history of science and technologies, and humanities in general. An approach along these lines is certainly the Dysonian SETI [20-23].

**Acknowledgements**. The authors wish to express their gratitude to an anonymous reviewer whose cogent comments and criticisms have improved a previous version of this manuscript. The authors also acknowledge financial support from the Ministry of Education, Science and Technological Development of the Republic of Serbia through the contract number 451-03-68/2020/14/20002, as well as projects #ON176021 ("Visible and Invisible Matter in Nearby Galaxies: Theory and Observations") and #ON179048. ("The Theory and Practice of Science in Society: Multidisciplinary, Educational, and Intergenerational Perspectives").

# References

1. G. Anglada-Escudé, P.J. Amado, J. Barnes, Z.M. Berdiñas, R.P. Butler, G.A. L. Coleman, I. de La Cueva, S. Dreizler, M. Endl, B. Giesers, S.V. Jeffers, J.S. Jenkins, H.R.A. Jones, M. Kiraga, M. Kürster, M.J. López-González, C.J. Marvin, N. Morales, J. Morin, R.P. Nelson,

J.L. Ortiz, A. Ofir, S.-J. Paardekooper, A. Reiners, E. Rodríguez, C. Rodríguez-López, L.F. Sarmiento, J.P. Strachan, Y. Tsapras, M. Tuomi and M. Zechmeister, "A terrestrial planet candidate in a temperate orbit around Proxima Centauri", *Nature*, **536**, pp. 437–440, 2016.

2. P.S. Muirhead, J.A. Johnson, K. Apps, J.A. Carter, T.D. Morton, D.C. Fabrycky, J.S. Pineda, M. Bottom, B. Rojas-Ayala, E. Schlawin, K. Hamren, K.R. Covey, J.R. Crepp, K.G. Stassun, J. Pepper, L. Hebb, E.N. Kirby, A. W. Howard, H.T. Isaacson, G.W. Marcy, D. Levitan, T. Diaz-Santos, L. Armus and J.P. Lloyd, "Characterizing the Cool KOIs. III. KOI 961: A Small Star with Large Proper Motion and Three Small Planets", *ApJ*, **747**, 144, 2012.

3. J.H. Steffen and W.M. Farr, "A Lack of Short-period Multiplanet Systems with Close-proximity Pairs and the Curious Case of Kepler-42", *ApJ*, **774**, L12, 2013.

4. M. Gillon, A.H.M.J. Triaud, B.-O. Demory, E. Jehin, E. Agol, K.M. Deck, S.M. Lederer, J. de Wit, A. Burdanov, J.G. Ingalls, E. Bolmont, J. Leconte, S.N. Raymond, F. Selsis, M. Turbet, K. Barkaoui, A. Burgasser, M.R. Burleigh, S.J. Carey, A. Chaushev, C.M. Copperwheat, L. Delrez, C.S. Fernand es, D.L. Holdsworth, E.J. Kotze, V. Van Grootel, Y. Almleaky, Z. Benkhaldoun, P. Magain and D. Queloz, "Seven temperate terrestrial planets around the nearby ultracool dwarf star TRAPPIST-1", *Nature*, **542**, pp. 456–460, 2017.

5. J.J. Bochanski, S. L. Hawley, K.R. Covey, A.A. West, I.N. Reid, D.A. Golimowski and Ž. Ivezić, "The Luminosity and Mass Functions of Low-mass Stars in the Galactic Disk. II. The Field", *AJ*, **139**, pp. 2679–2699, 2010.

6. G. Anglada-Escude, P. Arriagada, M. Tuomi, M. Zechmeister, J.S. Jenkins, A. Ofir, S. Dreizler, E. Gerlach, C.J. Marvin, A. Reiners, S.V. Jeffers, R.P. Butler, S.S. Vogt, P.J. Amado, C. Rodriguez-Lopez, Z.M. Berdinas, J. Morin, J.D. Crane, S.A. Shectman, I.B. Thompson, M. Diaz, E. Rivera, L.F. Sarmiento and H.R.A. Jones, "Two planets around Kapteyn's star: a cold and a temperate super-Earth orbiting the nearest halo red dwarf.", *MNRAS*, **443**, pp. L89–L93, 2014.

7. L.V. Ksanfomaliti, "Problem of habitation on planetary systems of red dwarf stars.", *JBIS*, **39**, pp. 416–417, 1986.

8. M.J. Heath, L.R. Doyle, M.M. Joshi and R.M. Haberle, "Habitability of Planets Around Red Dwarf Stars", *Origins of Life and Evolution of the Biosphere*, **29**, pp. 405–424, 1999.

9. J.C. Tarter, P.R. Backus, R.L. Mancinelli, J.M. Aurnou, D.E. Backman, G.S. Basri, A.P. Boss, A. Clarke, D. Deming, L.R. Doyle, E.D. Feigelson, F. Freund, D.H. Grinspoon, R.M. Haberle, I. Hauck, Steven A., M.J. Heath, T.J. Henry, J.L. Hollingsworth, M.M. Joshi, S.

Kilston, M.C. Liu, E. Meikle, I.N. Reid, L.J. Rothschild, J. Scalo, A. Segura, C.M. Tang, J.M. Tiedje, M.C. Turnbull, L.M. Walkowicz, A.L. Weber and R.E. Young, "A Reappraisal of The Habitability of Planets around M Dwarf Stars", *Astrobiology*, **7**, pp. 30–65, 2007.

10. A. Segura, L.M. Walkowicz, V. Meadows, J. Kasting and S. Hawley, "The Effect of a Strong Stellar Flare on the Atmospheric Chemistry of an Earth-like Planet Orbiting an M Dwarf", *Astrobiology*, **10**, pp. 751–771, 2010.

11. M. López-Morales, N. Gómez-Pérez and T. Ruedas, "Magnetic Fields in Earth-like Exoplanets and Implications for Habitability around M-dwarfs", *Origins of Life and Evolution of the Biosphere*, **41**, pp. 533–537, 2011.

12. A.W. Mann, E. Gaidos and M. Ansdell, "Spectro-thermometry of M Dwarfs and Their Candidate Planets: Too Hot, Too Cool, or Just Right?", *ApJ*, **779**, 188, 2013.

13. A.L. Shields, S. Ballard and J.A. Johnson, "The habitability of planets orbiting M-dwarf stars", *Phys. Rep.*, **663**, p. 1, 2016.

14. M. Lingam and A. Loeb, "Reduced Diversity of Life around Proxima Centauri and TRAPPIST-1", *ApJ*, **846**, L21, 2017.

15. R.J. Ritchie, A.W.D. Larkum and I. Ribas, "Could photosynthesis function on Proxima Centauri b?", *International Journal of Astrobiology*, **17**, pp. 147–176, 2018.

16. P. Behroozi and M.S. Peeples, "On the history and future of cosmic planet formation", *MNRAS*, **454**, pp. 1811–1817, 2015.

17. E. Zackrisson, P. Calissendorff, J. González, A. Benson, A. Johansen and M. Janson, "Terrestrial Planets across Space and Time", *ApJ*, **833**, 214, 2016.

18. J. Haqq-Misra, R.K. Kopparapu and E.T. Wolf, "Why do we find ourselves around a yellow star instead of a red star?", *International Journal of Astrobiology*, **17**, pp. 77–86, 2018.

19. G. Cocconi and P. Morrison, "Searching for Interstellar Communications", *Nature*, **184**, pp. 844–846, 1959.

20. J.T. Wright, B. Mullan, S. Sigurdsson and M.S. Povich, "The Ĝ Infrared Search for Extraterrestrial Civilizations with Large Energy Supplies. I. Background and Justification", *ApJ*, **792**, 26, 2014.

21. J.T. Wright, R.L. Griffith, S. Sigurdsson, M.S. Povich and B. Mullan, "The Ĝ Infrared Search for Extraterrestrial Civilizations with Large Energy Supplies. II. Framework, Strategy, and First Result", *ApJ*, **792**, 27, 2014.

22. E. Zackrisson, P. Calissendorff, S. Asadi and A. Nyholm, "Extragalactic SETI: The Tully-Fisher Relation as a Probe of Dysonian Astroengineering in Disk Galaxies", *ApJ*, **810**, 23, 2015.

23. R.J. Bradbury, M.M. Ćirković and G. Dvorsky, "Dysonian Approach to SETI: A Fruitful Middle Ground?", *JBIS*, **64**, pp. 156–165, 2011.

24. G. Gaylord Simpson, "The Nonprevalence of Humanoids", *Science*, **143**, pp. 769–775, 1964.

25. J. Diamond, *The Third Chimpanzee*, Hutchinson Radius, London, 1991.

26. S.J. Dick, "Cultural evolution, the postbiological universe and SETI", *International Journal of Astrobiology*, **2**, pp. 65–74, 2003.

27. M.M. Ćirković, *The Astrobiological Landscape: Philosophical Foundations of the Study of Cosmic Life*, Cambridge University Press, Cambridge, 2012.

28. J. Diamond, "Alone in a crowded universe", *Natural History*, **99**, pp. 30–34, reprinted in [53], 1990.

29. M.M. Ćirković, "Woodpeckers and Diamonds: Some Aspects of Evolutionary Convergence in Astrobiology", *Astrobiology*, **18**, pp. 491–502, 2018.

30. G.J. Vermeij, "Historical contingency and the purported uniqueness of evolutionary innovations", *Proceedings of the National Academy of Science*, **103**, pp. 1804–1809, 2006.

31. M.D. Papagiannis, "Natural Selection of Stellar Civilizations by the Limits of Growth", *QJRAS*, **25**, p. 309, 1984.

32. A.D.P. Hands, K.A. Ryden, N.P. Meredith, S.A. Glauert and R.B. Horne, "Radiation Effects on Satellites During Extreme Space Weather Events", *Space Weather*, **16**, pp. 1216–1226, 2018.

33. R. Pirjola, "Effects of space weather on high-latitude ground systems", *Advances in Space Research*, **36**, pp. 2231–2240, 2005.

34. J.B.L. Jones, R.D. Bentley, R. Hunter, R.H.A. Iles, G.C. Taylor and D.J. Thomas, "Space weather and commercial airlines", *Advances in Space Research*, **36**, pp. 2258–2267, 2005.

35. D.B. Contreira, F.S. Rodrigues, K. Makita, C.G.M. Brum, W. Gonzalez, N.B. Trivedi, M.R. da Silva and N.J. Schuch, "An experiment to study solar flare effects on radio-communication signals", *Advances in Space Research*, **36**, pp. 2455–2459, 2005.

36. S. Odenwald, J. Green and W. Taylor, "Forecasting the impact of an 1859-calibre super-storm on satellite resources", *Advances in Space Research*, **38**, pp. 280–297, 2006.

37. V. Bothmer and I. A. Daglis, *Space Weather – Physics and Effects*, Praxis Publishing and Springer Science+Business Media, 2007.

38. M. Güdel, "Stellar Radio Astronomy: Probing Stellar Atmospheres from Protostars to Giants", *ARA&A*, **40**, pp. 217–261, 2002.

39. A.A. West, K.L. Weisenburger, J. Irwin, Z.K. Berta-Thompson, D. Charbonneau, J. Dittmann and J.S. Pineda, "An Activity-Rotation Relationship and Kinematic Analysis of Nearby Mid-to-Late-Type M Dwarfs", *ApJ*, **812**, 3, 2015.

40. J.R.A. Davenport, D.M. Kipping, D. Sasselov, J.M. Matthews and C. Cameron, "MOST Observations of Our Nearest Neighbor: Flares on Proxima Centauri", *ApJ*, **829**, L31, 2016.

41. M. Lingam and A. Loeb, "Risks for Life on Habitable Planets from Superflares of Their Host Stars", *ApJ*, **848**, 41, 2017.

42. W.S. Howard, M.A. Tilley, H. Corbett, A. Youngblood, R.O.P. Loyd, J.K. Ratzloff, N.M. Law, O. Fors, D. del Ser, E. L. Shkolnik, C. Ziegler, E.E. Goeke, A.D. Pietraallo and J. Haislip, "The First Naked-eye Superflare Detected from Proxima Centauri", *ApJ*, **860**, L30, 2018.

43. J. Villadsen and G. Hallinan, "Ultra-wideband Detection of 22 Coherent Radio Bursts on M Dwarfs", *ApJ*, **871**, 214, 2019.

44. M.A. MacGregor, A.J. Weinberger, D.J. Wilner, A.F. Kowalski and S.R. Cranmer, "Detection of a Millimeter Flare from Proxima Centauri", *ApJ*, **855**, L2, 2018.

45. R.H. Gray and K.B. Marvel, "A VLA Search for the Ohio State "Wow"", *ApJ*, **546**, pp. 1171–1177, 2001.

46. M. Hippke, "Benchmarking information carriers", *Acta Astronautica*, **151**, pp. 53–62, 2018.

47. J.M. Cordes, J.W. Lazio and C. Sagan, "Scintillation-induced Intermittency in SETI", *ApJ*, **487**, pp. 782–808, 1997.

48. J.S. Pineda and G. Hallinan, "A Deep Radio Limit for the TRAPPIST-1 System", *ApJ*, **866**, 155, 2018.

49. P.D. Ward and D. Brownlee, *Rare Earth: Why Complex Life Is Uncommon in the Universe*, Springer, New York, 2000.

50. C.H. Lineweaver, *Paleontological Tests: Human-Like Intelligence Is Not a Convergent Feature of Evolution*, Springer, Dordrecht, 2008.

51. M.M. Ćirković and B. Vukotić, "Long-term prospects: Mitigation of supernova and gamma-ray burst threat to intelligent beings", *Acta Astronautica*, **129**, pp. 438–446, 2016.

52. C. Walters, R.A. Hoover and R.K. Kotra, "Interstellar colonization: A new parameter for the Drake equation?", *Icarus*, **41**, pp. 193–197, 1980.

53. B. Zuckerman and M.H. Hart (eds.) *Extraterrestrials. Where are they?*, Cambridge University Press, Cambridge, 1995.